\documentclass[aps, pra, twocolumn, superscriptaddress, amsmath,  tightenlines, longbibliography]{revtex4-2}

\usepackage{dcolumn}
\usepackage{graphicx}
\usepackage{epstopdf}
\usepackage{mathrsfs}
\usepackage{subfigure}
\usepackage{booktabs}
\usepackage{amsmath,amsfonts}
\usepackage{physics}
\usepackage{dsfont}
\usepackage{amstext}
\usepackage{amssymb}
\usepackage{amsbsy}
\usepackage{bbm}
\usepackage{amsthm}
\usepackage{graphicx}
\usepackage{xcolor}
\usepackage{CJK}
\usepackage[colorlinks,urlcolor=blue,linkcolor=blue,citecolor=blue]{hyperref}
\usepackage{soul}

\usepackage{url}
\usepackage[colorlinks]{hyperref}
\hypersetup{%
	plainpages=true,
	breaklinks=true,       
	hypertexnames=false,  
	pageanchor=true,
	colorlinks=true,
	linkcolor={blue},
	citecolor={blue},
	urlcolor={blue},
	anchorcolor={black}
}

 \makeatletter

\newcommand{\Rmnum}[1]{\expandafter\@slowromancap\romannumeral #1@}
\makeatother

\begin{document}

\title{Suppressing Degradation in Quantum Batteries by Electromagnetically-induced Transparency}

\author{Jin-Tian Zhang}
\address{School of Physics and Astronomy, Applied Optics Beijing Area Major
Laboratory, Beijing Normal University, Beijing 100875, China}
\address{Key Laboratory of Multiscale Spin Physics, Ministry of Education,
Beijing Normal University, Beijing 100875, China}

\author{Cheng-Ge Liu}
\address{School of Physics and Astronomy, Applied Optics Beijing Area Major
Laboratory, Beijing Normal University, Beijing 100875, China}
\address{Key Laboratory of Multiscale Spin Physics, Ministry of Education,
Beijing Normal University, Beijing 100875, China}

\author{Qing Ai}
\email{aiqing@bnu.edu.cn}
\address{School of Physics and Astronomy, Applied Optics Beijing Area Major
Laboratory, Beijing Normal University, Beijing 100875, China}
\address{Key Laboratory of Multiscale Spin Physics, Ministry of Education,
Beijing Normal University, Beijing 100875, China}

\begin{abstract}
Quantum batteries (QBs), as emerging quantum devices for energy storage and transfer, have attracted significant attention due to their potential to surpass classical batteries in charging efficiency and energy density. However, interactions between a QB and its environment result in decoherence, which significantly reduces its operational lifespan. In this work, we propose suppressing the aging of QB{s} by introducing the electromagnetically{-}induced transparency (EIT). Specifically, we model a four-level atom as a QB with an effective two-level system enabled by the EIT, while the photons in the cavity serve as the energy charger. By comparing the energy and extractable work of the QB with and without the EIT effect, we demonstrate that the QBs incorporating the EIT exhibit enhanced resistance to {spontaneous decay} as compared to their counterparts without the EIT. We believe that our findings may provide valuable insights and shed the light on the design principles for mitigating the {degradation} of the QBs.

\end{abstract}
\maketitle

\section{Introduction}

Classical batteries, as electrochemical devices, store energy and
provide power to electrical equipments \cite{Dunlap1999PRB,Castro2009RevModPhys,Parmananda1993PRE}. Whether it is the 150 kWh battery packs in electric vehicles or the dry cells in remote controls, batteries are ubiquitous in everyday life \cite{campaioli2018quantumbatteriesreview}. However, with rapid technological advancements, the demand for miniaturization has become increasingly urgent in cutting-edge fields such as quantum computing \cite{Tordrup2008PRL,Berthusen2025PRX,Pettersson2025PRX}, nanotechnology \cite{Soloviev2021PRA}, and quantum communication \cite{Vincent1997,Labont2024PRX}. This trend has driven the miniaturization of batteries{. W}hen their size is reduced to the atomic and molecular scales, quantum{-}mechanical effects become significant. This gave rise to the concept of quantum batteries (QBs) \cite{campaioli2018quantumbatteriesreview,Song2024PRL,Kamin2020PhysRevE,Le2018PRA}.

Quantum thermodynamics, {which} explores energy conversion,
heat transfer, and entropy evolution in quantum systems, seeks to
understand how quantum effects influence energy storage, transfer,
and conversion \cite{Sebastian2019book,Mayer2023CommunPhys,Maslennikov2019NatCommun,Zhou2024PRB,Bera2024PhysRevResearch}. One of its significant applications is the QB \cite{Campaioli2024RMP,Bhattacharjee2021EurPhysJB}.
Composed of quantum bits (qubits) or quantum oscillators, the QBs offer several advantages over traditional batteries, such as significantly faster charging times, higher energy-storage efficiency,
greater energy density, and increased precision and control \cite{Alicki2013PRE,Ferraro2018PRL}. These
features make QBs a promising solution to energy challenges
in advanced technologies. Since the concept was introduced in 2013 \cite{Alicki2013PRE}, the primary research { has been focused} on how to store and extract energy efficiently in QBs \cite{Andolina2019PRL}. Currently, the QB models based on two-level systems
are widely studied \cite{Pirmoradian2019PhysRevA,Crescente2020PRB,Guo2024PRA,Song2024PRL}. These batteries can interact with external {agents},
such as driving fields, thermal baths, or other two-level systems,
to achieve charging. Despite their potential, QBs face
practical challenges, one of the most notable being aging.
In QBs, aging refers to the gradual decline in energy
storage and extraction efficiency over time or after repeated charging{-}and{-}discharging cycles. Aging leads to reduced capacity, slower
energy extraction, and even an inability to store energy effectively.
In quantum systems, aging arises primarily from interactions with
the external environment, e.g. thermal baths, noise, or electromagnetic
fields \cite{Pirmoradian2019PhysRevA}, which disrupt quantum states and lead to the loss of quantum
coherence {and} entanglement \cite{Kamin2020PhysRevE}. This phenomenon,
known as decoherence, significantly impacts the energy storage and
extraction processes \cite{Andolina2019PRL}, leading to performance degradation or battery
aging. Decoherence is the process by which a quantum system loses
its quantum{-}state coherence over time. Numerous methods have been
proposed to suppress decoherence, such as environment
engineering \cite{2022XuPRA}, feedback control \cite{Yao2022PRE}, {and Floquet engineering \cite{Bai2020PRA}.} Inspired by these discoveries, in this paper, we explores using {the electromagnetically-induced transparency (EIT)} to suppress the aging of QBs.

{The EIT} is a quantum-optical
phenomenon widely used in optics and quantum information processing \cite{Fleischhauer2005RevModPhys,Boller1991PRL,Harris1990PRL}.
A classical EIT system consists of a three-level system with the states $|d\rangle$, $|m\rangle$, and $|e\rangle$ \cite{Paspalakis2001PRA}. Here,
$|d\rangle$ is the {initial} state, and $|m\rangle$ is the {target}
state, and $|e\rangle$ is the {intermediate} state. The probe field acts
between $|d\rangle$ and $|e\rangle$, while the driving field acts
between $|m\rangle$ and $|e\rangle$. Under these conditions, the system can enter a {coherent state}, which is superposition of the bright states and dark states. The bright state is
usually associated with the {intermediate} state $|e\rangle$, while
the dark state is typically a superposition of {$|d\rangle$ and $|m\rangle$} \cite{Wang2018PRA,Fleischhauer2000PRL,Reed2010PRL}.
Since the dark state does not involve the {intermediate} state $|e\rangle$,
it is unaffected by the dissipation due to the coupling {to} the environment and is thus highly stable. This stability makes dark states less sensitive to environmental noise,
and when a {quantum} system is in a dark state, energy loss can be avoided.
The aging of QBs is largely due to decoherence caused by interactions
with the external environment. This insight suggests that dark states
could be employed to suppress decoherence, thereby extending the operational
lifespan of QBs. Compared to other methods for suppressing
decoherence, {the} EIT offers high stability and is relatively easy to implement,
making it a viable solution to the problem of the QB aging. References~\cite{Quach2020PRA,2019SantosPRE} also introduce dark states to suppress the dissipation in QB. Unlike them, we use a four-level atom as the QB and, by adjusting the frequency of the driving light, we can enhance the ergodicity of the QB.

This paper is structured as follows. In the next section, we introduce the QB's model. By incorporating the EIT effect into a four-level atomic system, we can effectively obtain a two-level atom with the dark state as the excited state. The detailed derivations are outlined in Appendix~\ref{sec:AppendixA}. In Sec.~\ref{sec:Simulation}, we consider that the four-level atom is placed in a cavity {and} can be charged by the photons inside the cavity. We investigate the energy and ergotropy of the QB with different numbers of atoms and photons with and without the EIT effect. Some of the detailed calculations are provided in Appendix~\ref{sec:AppendixB}. We also analytically calculate the time evolution by the Wei-Norman algebra in Appendix~\ref{sec:AppendixC}. Our results show that the introduction of the EIT can effectively suppress the decay of the QB's energy. Finally, in the Sec.~\ref{sec:Conclusion}, we summarize our {main} findings.

\section{Model}
\label{sec:Model}

\begin{figure}
\includegraphics[width=8.5cm]{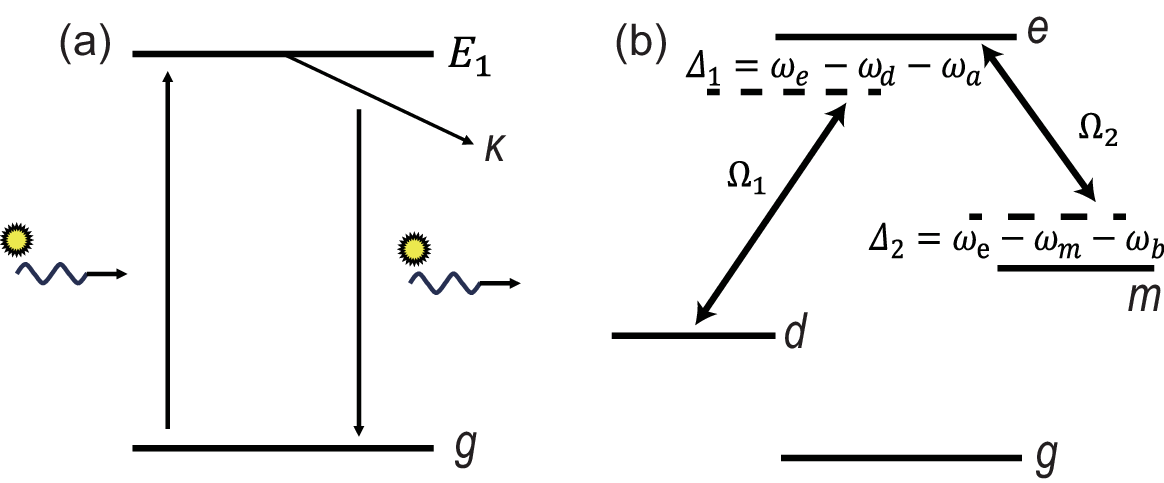}
\caption{Schematic of a QB against degradation by {the} EIT. (a) The QB is a two-level
system including the ground state $|g\rangle$ and the dark state
$|E_{1}\rangle$. When a photon is absorbed and the atom is excited
from $|g\rangle$ to $|E_{1}\rangle$, the QB is charged. However,
when a photon is emitted and the atom is deexcited from $|E_{1}\rangle$
to $|g\rangle$, the QB is discharged. (b) In order to realize the
dark state, a three-level configuration with two lower states $|d\rangle$
and $|m\rangle$, and the higher intermediate state $|e\rangle$ is
employed. The Rabi frequencies for the transitions $|d\rangle\leftrightarrow|e\rangle$
and $|m\rangle\leftrightarrow|e\rangle$ are respectively $\Omega_{1}$
and $\Omega_{2}$. And $\Delta_{1}$ and $\Delta_{2}$ are respectively
the two detunings.}\label{fig:EIT}
\end{figure}

{First of all, w}e consider a three-level system as shown in Fig.~\ref{fig:EIT}(b).
The Hamiltonian reads
\begin{eqnarray}
H & \!\!=\!\! & \omega_{d}\vert d\rangle\langle d\vert+\omega_{e}\vert e\rangle\langle e\vert+\omega_{m}\vert m\rangle\langle m\vert+2\Omega_{1}\cos\omega_{a}t\vert e\rangle\langle d\vert\nonumber \\
 & \!\!\!\! & +2\Omega_{2}\cos\omega_{b}t\vert e\rangle\langle m\vert+\mathrm{h.c.},
\end{eqnarray}
where $\omega_{d}$, $\omega_{e}$ and $\omega_{m}$ are respectively
the energies of the states $|d\rangle$, $|e\rangle$ and $|m\rangle$,
the Rabi frequencies of the transitions $|d\rangle\leftrightarrow|e\rangle$
and $|m\rangle\leftrightarrow|e\rangle$ are respectively $\Omega_{1}$
and $\Omega_{2}$, $\omega_{a}$ and $\omega_{b}$ are respectively
the driving frequencies. Here, we assume $\hbar=1$ for simplicity.

In the rotating frame with respect to $U=\exp[i(\omega_{a}|e\rangle\langle e|+\omega_{c}|m\rangle\langle m|)t]$,
where $\omega_{b}+\omega_{c}=\omega_{a}$. Additionally, we apply the rotating-wave approximation to eliminate the rapidly oscillating, high-frequency terms. The effective Hamiltonian $H_{\mathrm{eff}}=U^{\dagger}HU+i\dot{U^{\dagger}}U$
is simplified as
\begin{eqnarray}
H_{\mathrm{eff}} & = & \omega_{d}|d\rangle\langle d|+(\omega_{e}+\omega_{a})|e\rangle\langle e|+(\omega_{m}+\omega_{c})|m\rangle\langle m|\nonumber \\
 &  & +\Omega_{1}(|e\rangle\langle d|+|d\rangle\langle e|)+\Omega_{2}(|e\rangle\langle m|+|m\rangle\langle e|).
\end{eqnarray}
On account of the dissipation on $\vert d\rangle$ with decay rate
$\kappa$, by using the non-Hermitian Hamiltonian approach, the effectively
Hamiltonian can be written in the matrix form as
\begin{equation}
H_{\mathrm{eff}}^{\prime}=\left(\begin{array}{ccc}
\omega_{d}-i\kappa & \Omega_{1} & 0\\
\Omega_{1} & \omega_{e}+\omega_{a} & \Omega_{2}\\
0 & \Omega_{2} & \omega_{m}+\omega_{c}
\end{array}\right).
\end{equation}

After some algebra, the effective Hamiltonian can be rewritten as
\begin{align}
H_{\mathrm{eff}}^{\prime} & =H_{0}+I(\omega_{d}-i\kappa),
\end{align}
where
\begin{align}
H_{0} & =\left(\begin{array}{ccc}
0 & \Omega_{1} & 0\\
\Omega_{1} & \omega_{1} & \Omega_{2}\\
0 & \Omega_{2} & \omega_{2}
\end{array}\right),\\
\omega_{1} & =\omega_{e}+\omega_{a}-\omega_{d}+i\kappa,\\
\omega_{2} & =\omega_{m}+\omega_{c}-\omega_{d}+i\kappa.
\end{align}

According to Appendix~\ref{sec:AppendixA}, the Hamiltonian
$H_{0}$ can be diagonalized as
\begin{equation}
H_{0}=\sum_{j=1}^{3}x_{j}|E_{j}\rangle\langle E_{j}|,
\end{equation}
where the three eigen states are
\begin{align}
|E_{1}\rangle & \simeq\frac{\Omega_{2}}{N_{1}}(-\Omega_{2}|d\rangle+\Omega_{1}|m\rangle)\label{eq.E1},\\
|E_{2}\rangle & \simeq\frac{\Omega_{1}}{N_{2}}(\Omega_{1}|d\rangle+\Omega|e\rangle+\Omega_{2}|m\rangle),\\
|E_{3}\rangle & \simeq\frac{\Omega_{1}}{N_{3}}(\Omega_{1}|d\rangle-\Omega|e\rangle+\Omega_{2}|m\rangle).
\end{align}
By substituting $H_{0}$ into $H_{\mathrm{eff}}^{\prime}$, we can obtain
\begin{align}
H_{\mathrm{eff}}^{\prime} & =\sum_{j=1}^{3}x_{j}^{\prime}|E_{j}\rangle\langle E_{j}|,
\end{align}
where
\begin{eqnarray}
\!\!\!x_{1}^{\prime} & \!\!\!=\!\!\! & \frac{\Omega_{1}^{2}}{\Omega^{2}}(\omega_{m}+\omega_{c}-\omega_{d})+\omega_{d}-i\frac{\Omega_{2}^{2}}{\Omega^{2}}\kappa,\\
\!\!\!x_{2}^{\prime} & \!\!\!=\!\!\! & \Omega+\frac{\omega_{e}+\omega_{a}}{2}+\frac{\omega_{m}+\omega_{c}}{2\Omega^{2}}\Omega_{2}^{2}+\frac{\Omega_{1}^{2}\omega_{d}}{2\Omega^{2}}\!-\!\frac{i\Omega_{1}^{2}\kappa}{2\Omega^{2}},\\
\!\!\!x_{3}^{\prime} & \!\!\!=\!\!\! & -\Omega+\frac{\omega_{e}+\omega_{a}}{2}+\frac{\omega_{m}-\omega_{c}}{2\Omega^{2}}\Omega_{2}^{2}+\frac{\Omega_{1}^{2}\omega_{d}}{2\Omega^{2}}\!-\!\frac{i\Omega_{1}^{2}\kappa}{2\Omega^{2}}.
\end{eqnarray}

The imaginary parts of {$x_{j}$'s} represent the relaxation rate
of the system. Compared to the original decay rate $\kappa$, the
current decay rates are respectively {$\Omega_{2}^{2}\kappa/\Omega^{2}$ for $|E_1\rangle$
and $\Omega_{1}^{2}\kappa/2\Omega^{2}$ for $|E_2\rangle$ and $|E_3\rangle$}. If we tune $\Omega_{1}\gg\Omega_{2}$,
we have $\Omega_{2}^{2}\kappa/\Omega^{2}\simeq0$ and $\Omega_{1}^{2}\kappa/2\Omega^{2}\simeq\kappa/2$.
In other words, $|E_{1}\rangle$ is the dark state because its relaxation
has been significantly suppressed, while $|E_{2}\rangle$ and $|E_{3}\rangle$
are the two bright states. In this regards, we utilize the dark state
and the ground state to establish a QB against the aging.

\section{Numerical Simulation and Discussions}
\label{sec:Simulation}

\begin{figure}
\includegraphics[width= 7cm]{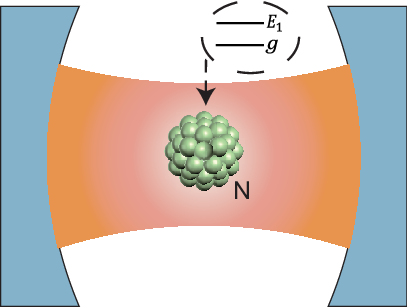}
\caption{Schematic of a QB with a collective of atoms in a cavity. There are
effectively $N$ two-level atoms {with the ground state $g$ and the dark
state $E_{1}$} located in a cavity. }\label{fig:Scheme}
\end{figure}

Based on the above discussions, as shown in Figure~\ref{fig:Scheme}, when $N=1${, we} consider a two-level
atom with the ground state $|g\rangle$ and the dark state $|E_{1}\rangle$
in a cavity. The total Hamiltonian is
\begin{equation}
H=x_{1}^{\prime}|E_{1}\rangle\langle E_{1}|+\omega a^{+}a+Ja^{+}\sigma^-+Ja\sigma^+,
\end{equation}
where \(J\) is the coupling constant between the atom and the cavity, 
\(a^\dagger\) (\(a\)) is the creation (annihilation) operator of the cavity mode with frequency \(\omega\). $\sigma^+=|E_{1}\rangle\langle g|=(\sigma^-)^\dagger$ is the atomic raising operator.

In the subspace spanned by the two bases $|n\rangle|E_{1}\rangle$
and $|n+1\rangle|g\rangle$, the total Hamiltonian can be rewritten
in the matrix form as
\begin{equation}
H=\left(\begin{array}{cc}
x_{1}^{\prime}-\frac{\omega}{2} & J\\
J & \frac{\omega}{2}
\end{array}\right)+\left(n+\frac{1}{2}\right)\omega,
\end{equation}
where the two eigen energies are
\begin{equation}
\lambda_{n}^{\pm}=\frac{1}{2}\left(x_{1}^{\prime}\pm\sqrt{4J^{2}+(\omega-x_{1}^{\prime})^{2}}\right)+\left(n+\frac{1}{2}\right)\omega,
\end{equation}
and the two eigen states are
\begin{align}
|\psi_{n}^{+}\rangle & =\frac{1}{N_{n}^{+}}\left(\begin{array}{c}
J\\
\lambda_{n}^{+}-x_{1}^{\prime}-n\omega
\end{array}\right),\label{eq:19}\\
|\psi_{n}^{-}\rangle & =\frac{1}{N_{n}^{-}}\left(\begin{array}{c}
J\\
\lambda_{n}^{-}-x_{1}^{\prime}-n\omega
\end{array}\right),\label{eq:20}\\
N_{n}^{\pm} & =\sqrt{\left(x_{1}^{\prime}+n\omega-\lambda_{n}^{\pm}\right)^{2}+J^{2}}
\end{align}
are the normalization constants.

Assuming the initial state $|n\rangle|E_1\rangle$, the time evolution
of the system can be given as
\begin{eqnarray}
|\psi(t)\rangle & = & \sum_{\alpha=\pm}\frac{\lambda_{n}^{\alpha}-x_{1}^{\prime}-n\omega}{N_{n}^{\alpha}}e^{-i\lambda_{n}^{\alpha}t}|\psi_{n}^{\alpha}\rangle.
\end{eqnarray}

\begin{figure}
\includegraphics[width=8.5cm]{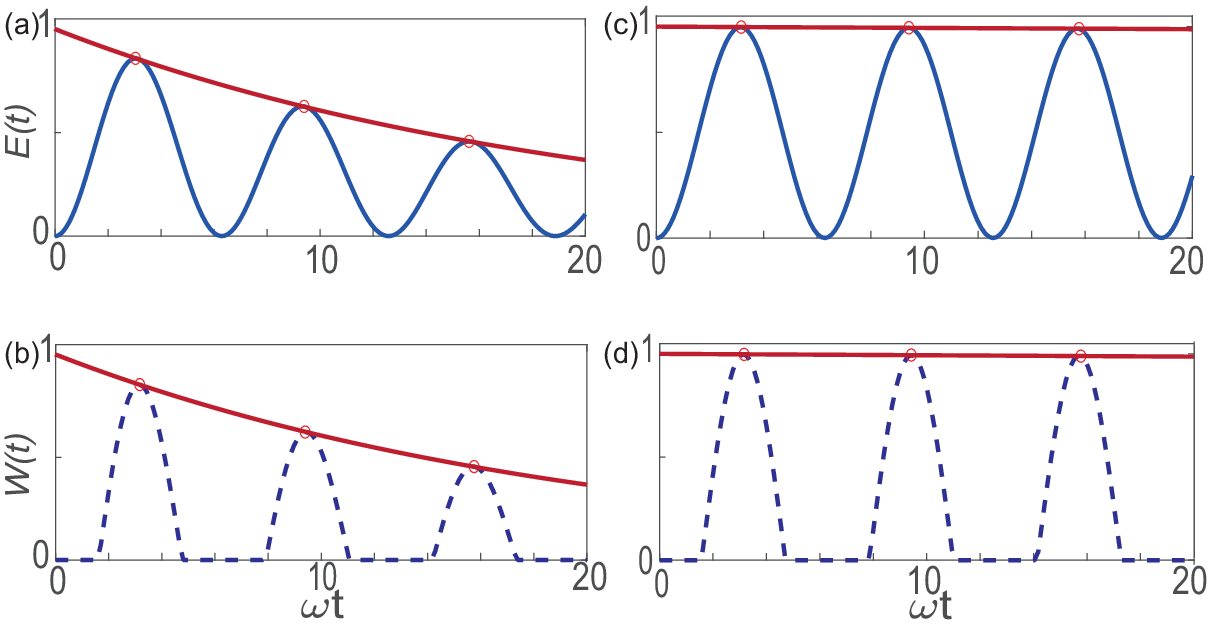}
\caption{The effects of the EIT on the system's energy $E(t)$ and ergodicity $W(t)$ with
$N=1$ atom and $n=1$ photon. (a)  $E(t)$ and (b) $W(t)$ without
the EIT, (c) $E(t)$ and (d) $W(t)$ with the EIT. The parameters
are $\omega=1$,\textcolor{red}{{} }$J=0.5\omega$, $\omega_{m}=0.5\omega$,
$\omega_{e}=\omega$, $\omega_{d}=0.25\omega$, $\Omega_{1}=50\omega$,
$\Omega_{2}=5\omega$, $\kappa=0.05\omega$. The red solid line
denotes the envelope which is numerically fitted by an exponential
decay.  }\label{fig:ergodicity}
\end{figure}

In the QB, the total Hamiltonian can be divided into three parts as
$H=H_{A}+H_{B}+H_{I}$, where $H_{A}$ is the Hamiltonian of the charger,
$H_{B}$ is the Hamiltonian of the QB, and $H_{I}$ is the interaction
Hamiltonian between them. The ergotropy is equal to {\cite{Andolina2019PRL}}
\begin{equation}
\begin{aligned}W(t) & =\mathrm{Tr}[\rho_{B}(t)H_{B}]-\mathrm{Tr}[\tilde{\rho}_{B}(t)H_{B}],\end{aligned}
\end{equation}
where the energy of the QB is $E(t)=\mathrm{Tr}[\rho_{B}(t)H_{B}]$,
$E_{0}(t)\mathrm{=Tr}[\tilde{\rho}_{B}(t)H_{B}]$ is the energy of
the passive state. Here, $\rho_{B}(t)=\sum_{n}r_{n}|r_{n}\rangle\langle r_{n}|$
is the reduced density matrix of the QB, $\tilde{\rho}_{B}(t){\equiv}\sum_{n}r_{n}|\varepsilon_{n}\rangle\langle\varepsilon_{n}|$,
$H_{B}=\sum_{n}\varepsilon_{n}|\varepsilon_{n}\rangle\langle\varepsilon_{n}|$,
$r_{1}\geq r_{2}\geq\cdots$, and $\varepsilon_{1}\leq\varepsilon_{2}\leq\cdots$. 
When $W(t)=0$, the system is in a passive
state, which implies that we can not extract any energy from the system.
Figure~\ref{fig:ergodicity} shows the system's ergodicity over time.
When the EIT is absent, we can see that $W(t)$ declines exponentially
with respect to the time. However, when the EIT is introduced, $W(t)$
generally oscillate with time. If we fit the envelope of $W(t)$ by
an exponential function, we could obtain $\exp(-5.01\times10^{-2}\omega t)$
and $\exp(-6.76\times10^{-4}\omega t)$ for the two cases respectively.
In other words, the decay of the ergodicity is reduced by 2 orders
of magnitude due to the presence of the EIT.

\begin{figure}
\includegraphics[width=8.5cm]{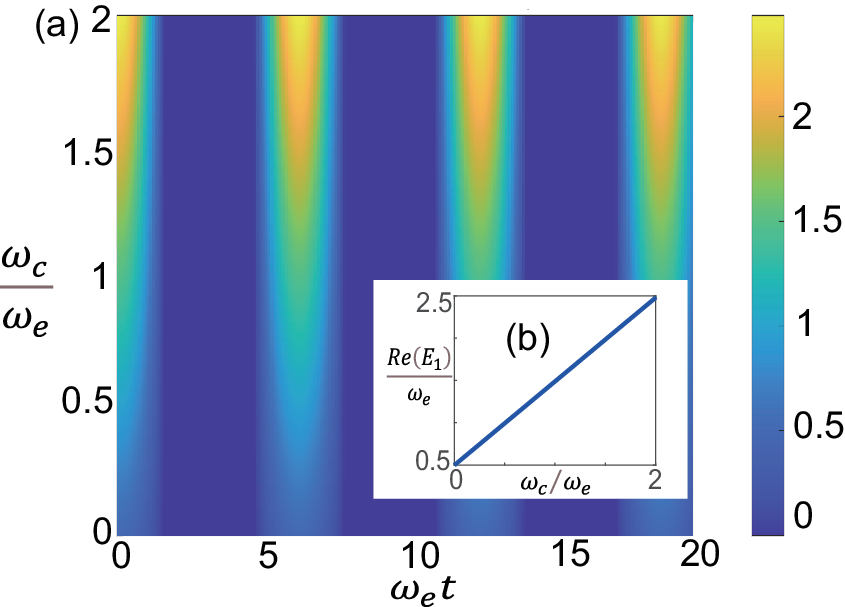}
\caption{ (a) The effects of the EIT on the system’s ergodicity as a function of the dimensionless time $\omega_e t$ and $\omega_c/\omega_e$ and (b) the inset shows the relation between $E_1$ and $\omega_c/\omega_e$. The parameters are $\omega_{e}=1$ $\Omega_{1}=50\omega_{e}$, $\Omega_{2}=5\omega_{e}$, $\Omega=\sqrt{\Omega_{1}^{2}+\Omega_{2}^{2}}$, $\omega_{d}=0.25\omega_{e}$, $\omega_{m}=0.5\omega_{e}$, $\omega_{a}=x\omega_{e}$, $\omega_{b}=0.5\omega_{e}$, $\omega_{c}=\omega_{a}-\omega_{b}$, $\kappa=0.05\omega_{e}$, $\omega=E_1$, and $J=0.5\omega_{e}$.}\label{fig omega}
\end{figure}

\begin{figure}
\includegraphics[width=8.5cm]{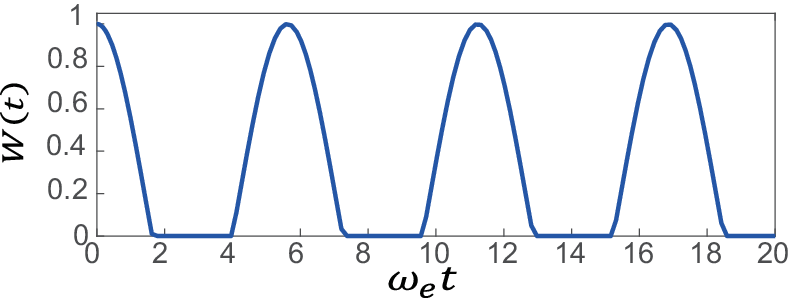}
\caption{Time evolution of the ergodicity
        $W(t)$ as a function of the dimensionless time 
        $\omega_{e} t$. The parameters are $\omega_{e}=1$ $E_1=\omega_{e}$, $J=0.5\omega_{e}$, and $\omega=0.5\omega_{e}$.}\label{fig10}
\end{figure}

After the EIT is introduced in QB, the dissipation can not only be reduced, but the ergodicity can also be enhanced by tuning the parameters. We construct a dark state $E_1$ as the effective excited state by coupling the states $d$ and $m$. According to Eq.~(\ref{eq.E1}), also shown in Fig.~\ref{fig omega}(b), the energy of the constructed excited state $E_1$ is positively correlated with $\omega_c$. Hence, by adjusting $\omega_c$, the energy of the excited state can be increased.

In Fig.~\ref{fig omega}, since $\omega_c = \omega_a - \omega_b$, we fix $\omega_b$ and vary $\omega_a$ to tune $\omega_c$. We found that the larger $\omega_c$ is, the greater the ergodicity becomes. By contrast, if we use the lower-dissipation state $e$ as the excited state, thereby forming a conventional two-level QB, cf. Fig.~\ref{fig10}, the maximum ergodicity only reaches 1. Compared to the traditional two-level QB, our system achieves a substantial improvement in ergodicity with the introduction of only two auxiliary states, while reducing the system's dissipation with the help of the EIT. As a concrete example, we propose constructing the QB using the $5S_{1/2}$, $5P_{1/2}$, $5P_{3/2}$, and $6S_{1/2}$ levels of rubidium atoms \cite{2005JitrikPAC}.

\begin{figure}
\includegraphics[width=8.5cm]{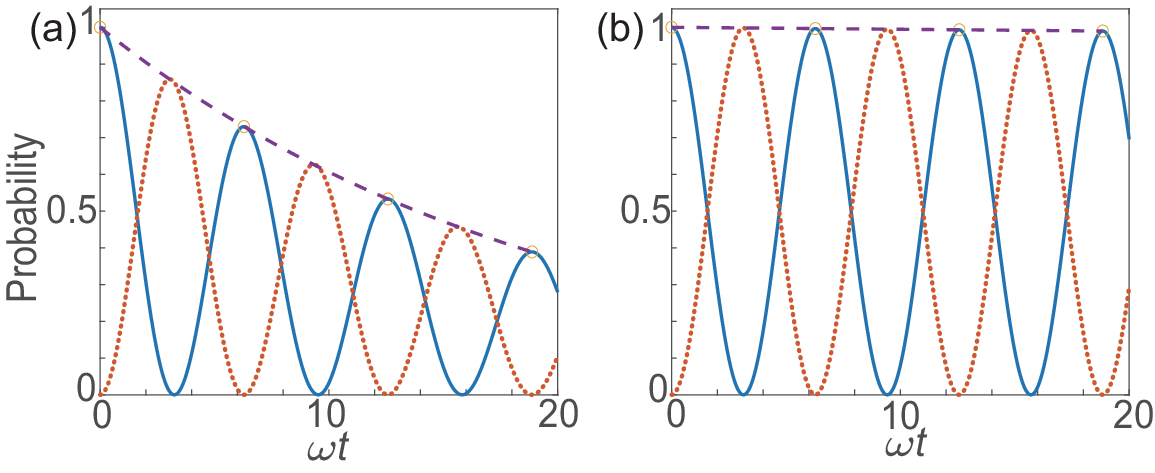}
\caption{The probabilities of the two states $|0\rangle|E_{1}\rangle$ (blue solid
line) and $|1\rangle|g\rangle$ (red {dotted} line) over time (a) without
the EIT and (b) with the EIT. The parameters are the same as Fig.~\ref{fig:ergodicity}. 
{The purple dashed line denotes the numerical fitting by an exponential function.}
}\label{fig:The=000020probabilities}

\end{figure}

Figure~ \ref{fig:The=000020probabilities} shows the quantum dynamic{al}
evolution of the system. Initially, the system is in $|0\rangle|E_{1}\rangle$.
In the absence of the EIT, the system decay significantly, as the
oscillatory behavior diminishes rapidly, leading to a noticeable reduction
in the amplitude of the probability curves. However, when the EIT
is introduced, the two probabilities oscillate with the same amplitude
and frequency. Notably, the oscillation{s occur} almost without any
observable decay. This suggests that the system's energy remains largely
unchanged. Similarly, we can also fit the envelope of the probabilities
by an exponential function. Thus, we obtain $\exp(-5.00\times10^{-2}\omega t)$
and $\exp(-5.38\times10^{-4}\omega t)$ for the two cases respectively, which
are consistent with those for the ergodicity in Fig.~\ref{fig:ergodicity}.
These suggest that due to the EIT, the dissipation has been significantly
suppressed and thus the ergodicity is improved.

{Furthermore, w}e consider a cavity with $N$ quasi-two-level atoms and a single
photon. The Hamiltonian reads
\begin{align}
H =\omega a^{\dagger}a+\sum_{j=1}^{N}\left(x_{1}^{\prime}\sigma_{j}^{+}\sigma_{j}^{-} +Ja\sigma_{j}^{+}+\mathrm{h.c.}\right),
\end{align}
where $\sigma_{j}^{+}=|E_{1}\rangle_{jj}\langle g|=(\sigma_{j}^{-})^{\dagger}$
is the raising operator of $j$th atom.

At time $t$, the total system is in the state{
\begin{align}
|\psi(t)\rangle & =\left[c_{0}(t)a^{\dagger}+\sum_{i=1}^{n}c_{i}(t)\sigma_{i}^{+}\right]|G\rangle|0\rangle,\label{eq:24}
\end{align}
where $|G\rangle=|g\rangle_{1}\otimes|g\rangle_{2}\otimes\cdots\otimes|g\rangle_{n}$ is the state for all atoms being in the ground states.}
By substituting Eq.~(\ref{eq:24}) into the Schr\"{o}dinger equation,
we can obtain
\begin{align}
i\dot{c_{0}}(t) & =\omega c_{0}(t)+J\sum_{j=1}^{n}c_{j}(t),\label{eq:25}\\
i\dot{c_{j}}(t) & =x_{1}^{\prime}c_{j}(t)+Jc_{0}(t),\label{eq:26}
\end{align}
where $c_{0}(0)=1$ and $c_{j}(0)=0$ are the initial conditions of
this system. By performing the Laplace transform $\mathcal{L}\left\{ df(t)/dt\right\} =s\tilde{f}(s)-f(0)$
on Eqs.~(\ref{eq:25}) and (\ref{eq:26}), we can obtain
\begin{align}
\tilde{c}_{0}(s) & =\frac{i(is-x_{1}^{\prime})}{(is-\omega_{0})(is-x_{1}^{\prime})-nJ^{2}},\\
\tilde{c}_{j}(s) & =\frac{iJ}{(is-\omega_{0})(is-x_{1}^{\prime})-nJ^{2}}.
\end{align}
After some algebra, $\tilde{c}_{0}(s)$ and $\tilde{c}_{j}(s)$ can be rewritten as
\begin{align}
\tilde{c}_{0}(s) & =\frac{i(is-x_{1}^{\prime})}{\left(s-s_{+}\right)\left(s-s_{-}\right)}\nonumber \\
 & =\frac{A_{+}}{s-s_{+}}+\frac{A_{-}}{s-s_{-}},\\
\tilde{c}_{j}(s) & =\frac{iJ}{\left(s-s_{+}\right)\left(s-s_{-}\right)}\nonumber \\
 & =\frac{B_{+}}{s-s_{+}}+\frac{B_{-}}{s-s_{-}},
\end{align}
where
\begin{align}
s_{\pm} & =-i\frac{(\omega_{0}+x_{1}^{\prime})\pm\sqrt{(\omega_{0}-x_{1}^{\prime})^{2}-nJ^{2}}}{2},\\
A_{\pm} & =\pm\frac{i(is_{\pm}-x_{1}^{\prime})}{s_{+}-s_{-}},\\
B_{\pm} & =\pm\frac{iJ}{s_{+}-s_{-}}.
\end{align}
By using the inverse Laplace transform, we can obtain
\begin{align}
c_{0}(t) & =A_{+}e^{s_{+}t}+A_{-}e^{s_{-}t},\\
c_{j}(t) & =B_{+}e^{s_{+}t}+B_{-}e^{s_{-}t}.
\end{align}
By partially tracing the degrees of freedom of the cavity, the reduced
density matrix of the atoms reads
\begin{eqnarray}
\rho_{B}(t) & = & |c_{0}|^{2}|G\rangle\langle G| +\sum_{j=1}^{n}|c_{j}|^{2}\sigma_{j}^{+}|G\rangle\langle G|\sigma_{j}^{-}\nonumber \\
 && +\sum_{j=1}^{n}\sum_{k\neq j}c_{j}c_{k}^{*}\sigma_{j}^{+}|G\rangle\langle G|\sigma_{k}^{-}.
\end{eqnarray}
As a result, the energy of the system $E_{B}^{N}(t)=\mathrm{Tr}[\rho_{B}(t)H_{B}]$
is equal to $x_{1}^{\prime}\sum_{j=1}^{n}|c_{j}(t)|^{2}$.

\begin{figure}
\includegraphics[width=8.5cm]{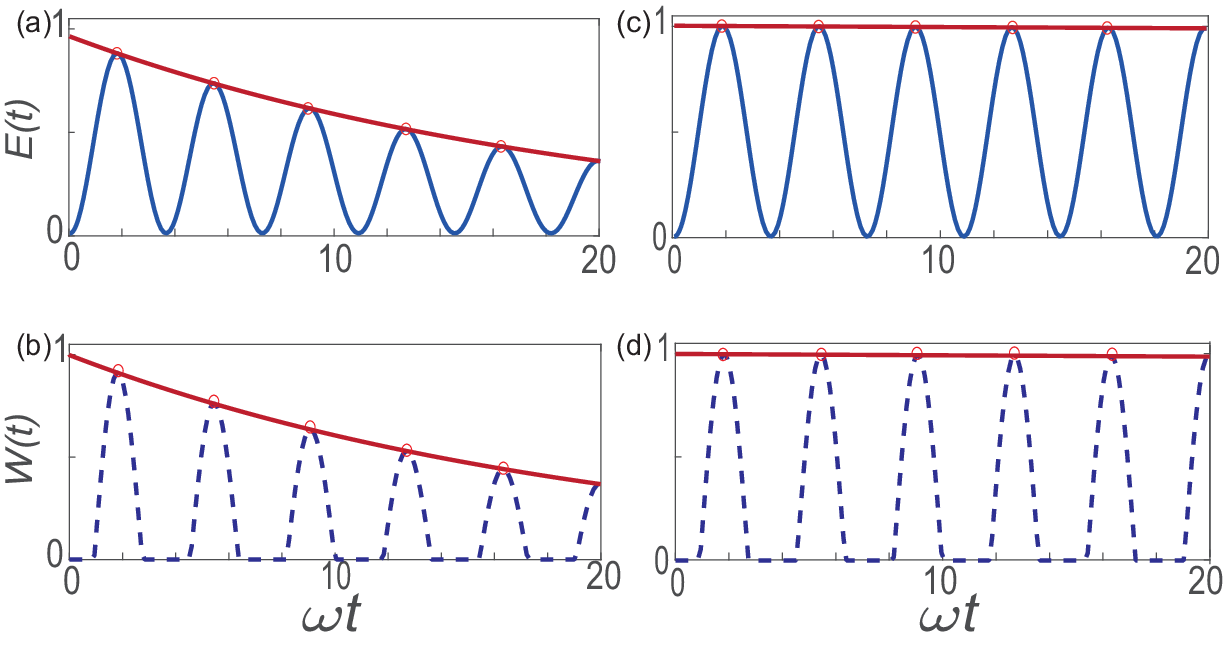}
\caption{The effects of the EIT on the system's energy $E(t)$ and ergodicity $W(t)$ with
$N=3$ atoms and $n=1$ photon. (a) $E(t)$ and (b) $W(t)$ without
the EIT, (c) $E(t)$ and (d) $E(t)$ with the EIT. The parameters
are $\omega=1$, $J=0.5\omega$, $\omega_{m}=0.5\omega$,
$\omega_{e}=\omega$, $\omega_{d}=0.25\omega$, $\Omega_{1}=50\omega$,
$\Omega_{2}=5\omega$, and $\kappa=0.05\omega$. The
red solid line denotes the envelope which is numerically fitted by
an exponential decay.}\label{fig:N3n1}
\end{figure}

Figure~\ref{fig:N3n1} shows the time evolution of the energy of
the system. Initially, the system is in an excited state.
The energy of the system decays significantly when {the} EIT is absent.
However, when {the} EIT is introduced, the energy decay is markedly suppressed.
After fitting the envelopes of the probabilities by an exponential
function, we can obtain $\exp(-5.03\times10^{-2}\omega t)$ and 
$\exp(-5.78\times10^{-4}\omega t)$
for the two cases, respectively. {Here}, we extend the single-atom
system to a multi-atom system. The results indicate that {the} EIT continues
to suppresses the decay of system energy, prolonging the lifetime
of QB.

{In the above calculations, we obtain the analytical solution by the Laplace transform. Alternatively, hereafter we will obtain the analytical solution by the Wei-Norman algebra.}
By introducing the collective operators
\begin{eqnarray}
J_{z} & = & \frac{1}{2}\sum_{j}\sigma_{j}^{z},\\
J_{+} & = & J_{-}^\dagger= \sum_{i}\sigma_{j}^{+},
\end{eqnarray}
and using the Holstein-Primakoff transformation \cite{Holstein1940PR}
\begin{eqnarray}
b^{\dagger}b & = & J_{z}+\frac{N}{2},\\
J_{+} & = & b^{\dagger}\sqrt{N}\sqrt{1-\frac{b^{\dagger}b}{N}}\simeq b^{\dagger}\sqrt{N},
\end{eqnarray}
the Hamiltonian can be rewritten as
\begin{equation}
H\simeq\omega a^{\dagger}a+x_{1}^{\prime}b^{\dagger}b+J_{N}(ab^{\dagger}+a^{\dagger}b),
\end{equation}
where the interaction $J_{N}=J\sqrt{N}$ between the cavity mode and the collective excitation
of the atoms has been enhanced by a factor $\sqrt{N}$.

Assuming that the initial state of the cavity is in the coherent state, and all of the two-level atoms are initially in the ground state, i.e., $|\Psi(0)\rangle=|\sqrt{N}\rangle_{A}\otimes|0\rangle_{B}$. 
According to Appendix~\ref{sec:AppendixC}, the state of the total system at time $t$ reads
\begin{eqnarray}
|\Psi(t)\rangle & \!\!=\!\! & |\sqrt{N}\cos(J_{N}t)\rangle_{A}\otimes|-i\sqrt{N}\sin(J_{N}t)\rangle_{B}.
\end{eqnarray}

\begin{figure}
\includegraphics[width=8.5cm]{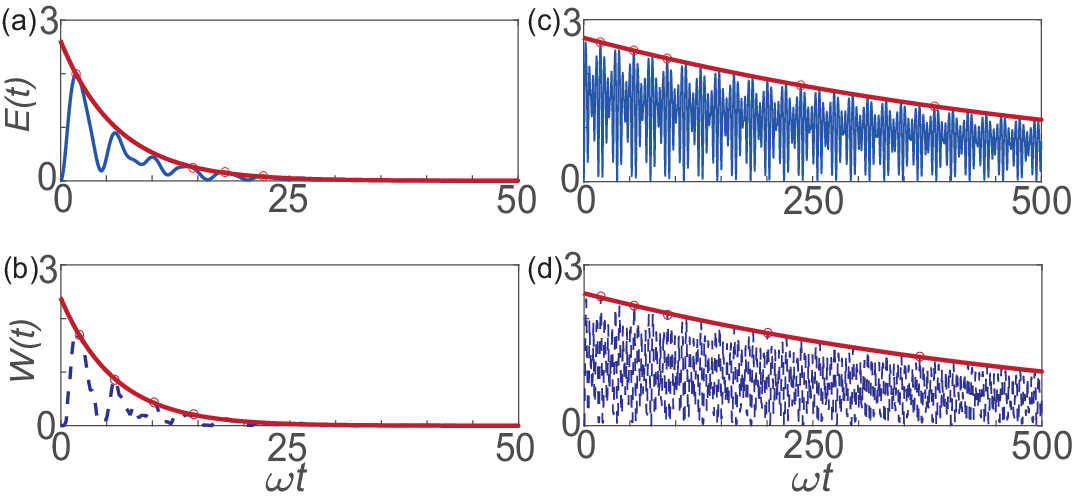}
\caption{The effects of the EIT on the system's energy $E(t)$ and ergodicity $W(t)$ with $N=3$ atoms and $n=3$ photons. (a) $E(t)$ and (b) $W(t)$ without
the EIT, (c) $E(t)$ and (d) $W(t)$ with the EIT. The parameters
are $\omega=1$, $J=0.5\omega$, $\omega_{m}=0.5\omega$,
$\omega_{e}=\omega$, $\omega_{d}=0.25\omega$, $\Omega_{1}=50\omega$,
$\Omega_{2}=5\omega$, $\kappa=0.05\omega$. The red solid line
denotes the enelope which is numerically fitted by an exponential
decay.}\label{fig:N3n3}
\end{figure}

When there are $3$ atoms and $3$ photons in the cavity, in Fig.~\ref{fig:N3n3},
we find that the maximum of the total energy is increased as compared
to the case with $3$ atoms and $1$ photon because more atoms can
be excited by photons. In addition, by fitting the envelopes with
an exponential function, we obtain a decay rate with {$-1.91\times10^{-3}\omega$
($-1.94\times10^{-1}\omega$) for the case with (without) the EIT.}
When the EIT is absent, both the system's energy and ergotropy exhibit
a rapid decay. However, when the EIT is introduced, both the energy
and ergotropy curves exhibit pronounced oscillatory behavior. This
indicates that the EIT does not only slow down the energy loss but
also preserves the oscillatory characteristics of the system.

When 2 photons and $N_a=2$ atoms in the cavity, on account of the dipole–dipole interaction \cite{2002FicekPR}, the Hamiltonian becomes
\begin{align}
H &= \omega\,a^{\dagger}a
  + H_{B}
  + J\sum_{i=1}^{N_a}\bigl(a\,\sigma_{i}^{+} + a^{\dagger}\,\sigma_{i}^{-}\bigr),
\end{align}
where the Hamiltonian of the QB reads
\begin{equation}
H_{B} 
= x_{1}^{\prime}\sum_{i=1}^{N_a}\sigma_{i}^{+}\sigma_{i}^{-}
  + \sum_{i\neq j}^{N_a}V\bigl(\sigma_{i}^{+}\sigma_{j}^{-} + \sigma_{i}^{-}\sigma_{j}^{+}\bigr).
\end{equation}
Its eigenvalues and eigenstates are respectively
\begin{align}
  E_{G} &= 0, 
    &\lvert G \rangle &= \lvert gg \rangle,\\
  E_{q} &= E_{1} + V, 
    &\lvert q \rangle &= \tfrac{1}{\sqrt{2}}\bigl(\lvert eg \rangle + \lvert ge \rangle\bigr),\\
  E_{p} &= E_{1} - V, 
    &\lvert p \rangle &= \tfrac{1}{\sqrt{2}}\bigl(\lvert eg \rangle - \lvert ge \rangle\bigr),\\
  E_{E} &= 2E_{1}, 
    &\lvert E \rangle &= \lvert ee \rangle.
\end{align}
Here, \(\lvert q \rangle\) is the symmetric state and \(\lvert p \rangle\) the antisymmetric state. Since
\begin{align}
  \sum_{i=1}^{2}\sigma_{i}^{+}\lvert G \rangle
    &= \lvert eg \rangle + \lvert ge \rangle = \sqrt{2}\,\lvert q \rangle,  \\
  \sum_{i=1}^{2}\sigma_{i}^{+}\lvert q \rangle
    &= \tfrac{1}{\sqrt{2}}(\lvert ee \rangle + \lvert ee \rangle)
    = \sqrt{2}\,\lvert E \rangle,  \\
  \sum_{i=1}^{2}\sigma_{i}^{+}\lvert p \rangle
    &= \tfrac{1}{\sqrt{2}}(\lvert ee \rangle - \lvert ee \rangle)
    = 0,  \\
  \sum_{i=1}^{2}\sigma_{i}^{+}\lvert E \rangle
    &= 0, 
\end{align}
in the eigenbasis of $H_B$, the total Hamiltonian can be rewritten as
\begin{equation}
H \!=\! \omega\,a^{\dagger}a
  + H_B
  + \sqrt{2}\,Ja\left(
      \,\lvert q \rangle\langle G\rvert
    + \,\lvert E \rangle\langle q\rvert\right)
    + \textrm{h.c.}
\end{equation}
Originally, the system include two excitation pathways,
\(\lvert gg \rangle \!\leftrightarrow\! \lvert eg \rangle \!\leftrightarrow\! \lvert ee \rangle\)
and
\(\lvert gg \rangle \!\leftrightarrow\! \lvert ge \rangle \!\leftrightarrow\! \lvert ee \rangle\).
However, after including the dipole–dipole interaction, only one pathway remains, i.e.,
\(\lvert gg \rangle \!\leftrightarrow\! \lvert q \rangle \!\leftrightarrow\! \lvert ee \rangle\), indicating a reduction in available transition channels.

Furthermore, we consider the effect of the dipole–dipole coupling on the ergodicity.  The QB Hamiltonian can be written in the basis \(\{\lvert G\rangle,\lvert p\rangle,\lvert q\rangle,\lvert E\rangle\}\) as
\begin{equation}
H_{B} =
\begin{pmatrix}
  0          & 0           & 0           & 0 \\
  0          & E_{1}-V     & 0           & 0 \\
  0          & 0           & E_{1}+V     & 0 \\
  0          & 0           & 0           & 2E_{1}
\end{pmatrix}.
\end{equation}
And the density matrix is
\(\rho_{B} = \sum_{\alpha=G,q,p,E}P_{\alpha}\lvert \alpha \rangle\langle \alpha \rvert\).
Ordering its eigenvalues \(P_{\alpha'}\) in descending order yields
\(\widetilde{\rho}_{B} = \sum_{\alpha'=1}^{4}P_{\alpha'}\lvert \alpha' \rangle\langle \alpha' \rvert\),
which will be used in evaluating the ergodicity.

\begin{figure}
\includegraphics[width=8.5cm]{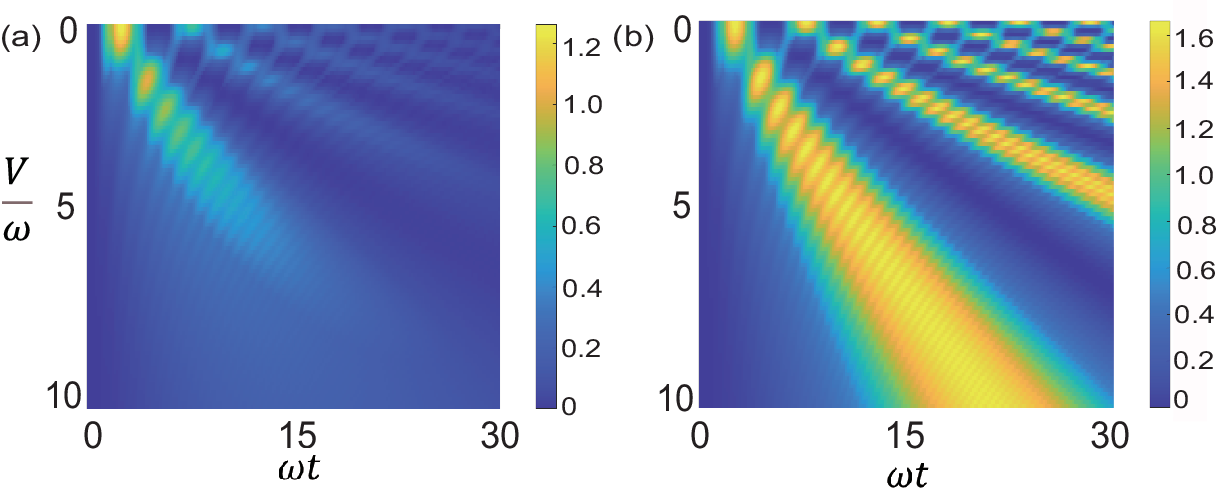}
\caption{Ergodicity as a function of the dipole-dipole interaction $V$ between the atoms and the time $t$ for 2 atoms and 2 photons in the cavity (a) without the EIT and (b) with the EIT. The parameters
are $\omega=1$, $J=0.5\omega$, $\omega_{m}=0.5\omega$,
$\omega_{e}=\omega$, $\omega_{d}=0.25\omega$, $\Omega_{1}=50\omega$,
$\Omega_{2}=5\omega$, and $\kappa=0.05\omega$.}\label{fig:9}
\end{figure}

When there are two atoms and two photons, the dynamics of the ergodicity is shown in Fig.~\ref{fig:9}. When the system does not include the EIT, the atoms exhibit large dissipation, causing the energy of the QB to decay more rapidly over time, as shown in Fig.~\ref{fig:9}(a). However, one can see that when the EIT is introduced, the decay of the maximum ergodicity of the QB will be significantly slowed down. Interestingly, the first time for the ergodicity to reach the maximum will be delayed and its duration will be prolonged as $V$ increases. 



Furthermore, we consider there are $N_a=3$ atoms in the cavity.
Since the antisymmetric states do not couple to the cavity field, we restrict ourselves to the four fully-symmetric eigenstates. They and their eigenvalues are listed as
\begin{align}
E_{0}\!\!&=\!\!0,~
  |D_{3,0}\rangle\!\!=\!\!|ggg\rangle,\\
E_{1}\!\!&=\!\!E_{1}+2V,~
  |D_{3,1}\rangle\!\!=\!\!\frac{1}{\sqrt{3}}\bigl(|egg\rangle+|geg\rangle+|gge\rangle\bigr),\\
E_{2}\!\!&=\!\!2E_{1}+2V,~
  |D_{3,2}\rangle\!\!=\!\!\frac{1}{\sqrt{3}}\bigl(|eeg\rangle+|ege\rangle+|gee\rangle\bigr),\\
E_{3}\!\!&=\!\!3E_{1},~
  |D_{3,3}\rangle\!\!=\!\!|eee\rangle.
\end{align}
Expanding the full Hamiltonian in the basis $\{|D_{3,k}\rangle\}$, one finds
\begin{align}
H
&=\omega a^{\dagger}a
+\sum_{k=0}^{3}E_{k}\,|D_{3,k}\rangle\langle D_{3,k}|\\
&\quad
+J\Bigl(
  \sqrt{3}\,a\,|D_{3,1}\rangle\langle D_{3,0}|
+2\,a\,|D_{3,2}\rangle\langle D_{3,1}|\\
&\quad
+\sqrt{3}\,a\,|D_{3,3}\rangle\langle D_{3,2}|
\Bigr)+\mathrm{h.c.}
\end{align}

\begin{figure}
\includegraphics[width=8.5cm]{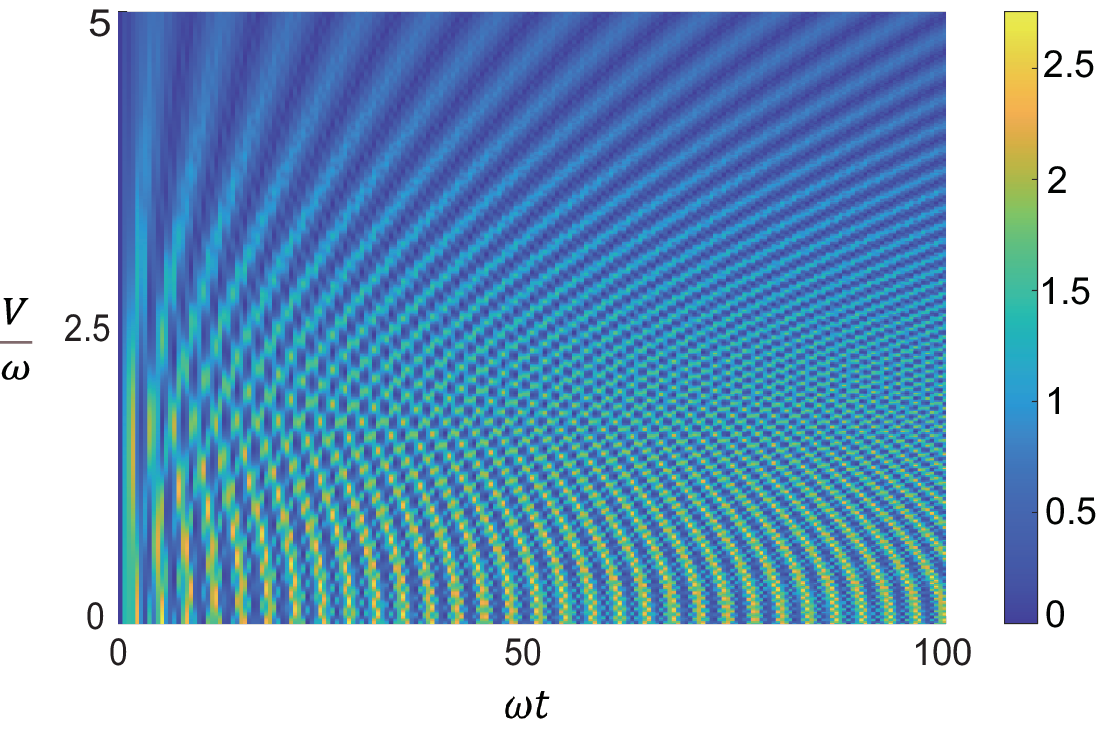}
\caption{Ergodicity as a function of \(V\) and \(t\) for a system with 3 atoms and 3 photons in the cavity with the EIT. The parameters
are $\omega=1$, $J=0.5\omega$, $\omega_{m}=0.5\omega$,
$\omega_{e}=\omega$, $\omega_{d}=0.25\omega$, $\Omega_{1}=50\omega$,
$\Omega_{2}=5\omega$, and $\kappa=0.05\omega$.}\label{fig:10}
\end{figure}
Unlike the case of two atoms in the cavity, when there are three atoms in the cavity, the level spacings between the two neighbouring eigenstates are respectively $E_{1}-E_{0}=E_{1}+2V$, $E_{2}-E_{1}=E_{1}$, $E_{3}-E_{2}=E_{1}-2V$. When the photon energy equals $E_{1}$, the detunings are $-2V$, $0$, and $2V$. Because the correspond oscillation frequencies are not the same, it leads to a more-complex situation. In Fig.~\ref{fig:10}, we find that as $V$ increases, the maximum ergodicity gradually decreases. Since the value of $V$ decreases rapidly with increasing interatomic distance, in practice we can reduce the influence of dipole-dipole interactions by increasing the distances between atoms.

When an electromagnetic drive with frequency $\omega_{L}$ and Rabi frequency \(\Omega\) is applied, and the cavity contains two photons, the total Hamiltonian with $N_a=2$ atoms becomes
\begin{equation}
H_{\mathrm{tot}}
=  x_{1}^{\prime}\sum_{i=1}^{N_a}\sigma_{i}^{+}\sigma_{i}^{-}
  + \sum_{i\neq j}V\bigl(\sigma_{i}^{+}\sigma_{j}^{-} + \sigma_{i}^{-}\sigma_{j}^{+}\bigr)
  + H_{\mathrm{d}},
\end{equation}
where
\begin{equation}
H_{\mathrm{d}}
= \sum_{i=1}^{N_a}\frac{\Omega}{2}\bigl(\sigma_{i}^{+}e^{-i\omega_{L}t}
  + \sigma_{i}^{-}e^{+i\omega_{L}t}\bigr).
\end{equation}
Transforming into the rotating frame defined by
$U(t)=\exp\!\Bigl(-i\,\omega_{L}\sum_{i}\sigma_{i}^{+}\sigma_{i}^{-}\,t\Bigr)$
and applying the rotating–wave approximation yields
\begin{equation}
H_{\mathrm{eff}}
=\sum_{i=1}^{N_a}\left(\Delta_{L}\sigma_{i}^{+}\sigma_{i}^{-}+\frac{\Omega}{2}\sigma_{i}^{+}\right) + \sum_{i\neq j}^{N_a}V\sigma_{i}^{+}\sigma_{j}^{-} +\textrm{h.c.},
\end{equation}
with \(\Delta_{L}=E_{1}-\omega_{L}\).  Focusing on the QB, we have
\begin{equation}
H_{B} = \Delta_{L}\,\sigma^{+}\sigma^{-}
      + \tfrac{\Omega}{2}(\sigma^{+} + \sigma^{-}),
\end{equation}
whose eigenvalues are
\(\displaystyle E_{\pm}=\bigl(\Delta_{L}\pm\sqrt{\Delta_{L}^{2}+\Omega^{2}}\bigr)/2\).
For \(\Delta_{L}\gg\Omega\), since
\(\sqrt{\Delta_{L}^{2}+\Omega^{2}}\approx \Delta_{L}+(\Omega^{2}/2\Delta_{L})\),
we have
\begin{align}
  E_{+}
  &= \Delta_{L} + \tfrac{\Omega^{2}}{4\Delta_{L}},\\
  E_{-}
  &= -\tfrac{\Omega^{2}}{4\Delta_{L}}.
\end{align}
In other words, the excited level is shifted up by \(\tfrac{\Omega^{2}}{4\Delta_{L}}\) while the ground level is shifted down by the same amount.  Consequently, the total Hamiltonian can be recast as
\begin{equation}
H_{\mathrm{tot}}
=  \Bigl(E_{1} + \tfrac{\Omega^{2}}{2\Delta_{L}}\Bigr)
    \sum_{i=1}^{N_a}\sigma_{i}^{+}\sigma_{i}^{-}
  + \sum_{i\neq j}V\sigma_{i}^{+}\sigma_{j}^{-} + \textrm{h.c.}
\end{equation}

\begin{figure}
\includegraphics[width=8.5cm]{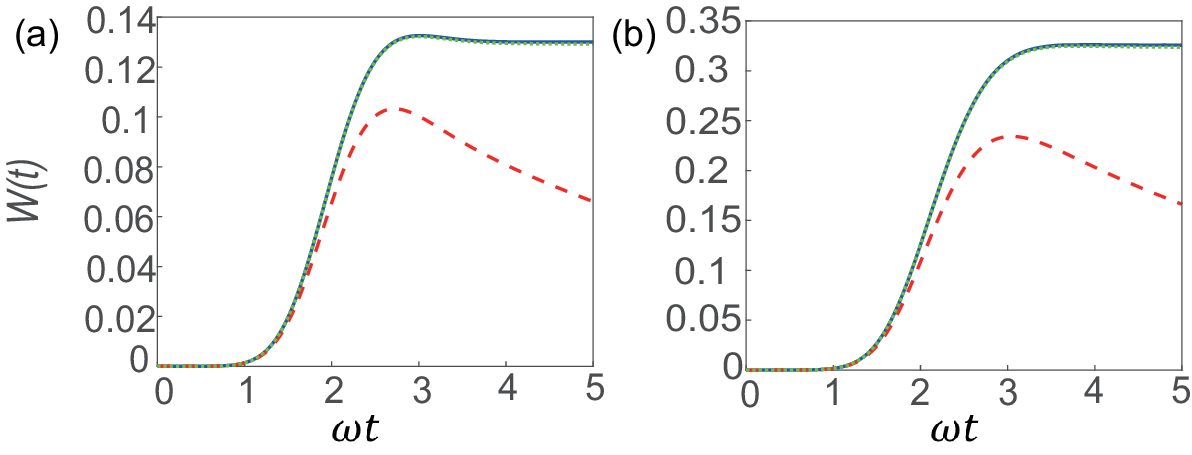}
\caption{The time evolution of the ergodicity with (a) $V=2$ and (b) $V=1$. The blue solid line represents the QB without dissipation, the green dotted line represents the QB with the EIT, and the red dashed line represents the QB without the EIT. The parameters
are $\omega=1$, $J=0.5\omega$, $\omega_{m}=0.5\omega$,
$\omega_{e}=\omega$, $\omega_{d}=0.25\omega$, $\Omega_{1}=50\omega$,
$\Omega_{2}=5\omega$, $t_{c}=1.6/\omega$, $\sigma=0.8/\omega$ and $\kappa=0.05\omega$. Notice that the blue solid line almost overlaps with the green dotted line.}\label{fig:11}
\end{figure}

We further consider a Gaussian pulse with $\Omega(t) = \Omega_{0} \exp\left[-(t - t_{c})^{2}/2\sigma^{2}\right]$ is used to excite the atoms, where $t_c$ is the center of the pulse, $\sigma$ is the width of the pulse. As shown in Fig.~\ref{fig:11}, the introduction of the EIT effectively suppresses the dissipation of the QB. In this process, the parameter $V$ also affects the ergodicity. Therefore, the dipole-dipole interaction should be minimized as much as possible.

\section{Implementation of a Four-Level Quantum Battery}

As a concrete example, we propose constructing the QB using the $5S_{1/2}$, $5P_{1/2}$, $5P_{3/2}$, and $6S_{1/2}$ levels of rubidium atoms \cite{2005JitrikPAC}. First of all, we prepare the rubidium atoms by heating a Rb dispenser to release vapor into the ultra-high-vacuum chamber. We then employ a magneto-optical trap (MOT) \cite{2024ZengPRL,2022GaudesiusPRA,2002SchoserPRA}, using counter-propagating $\sigma^+/\sigma^-$ cooling beams red-detuned by 10–20\,MHz and an anti-Helmholtz coil generating a position-dependent magnetic field gradient, to cool the atoms from thermal velocities down to approximately 100\,$\mu$K. 

Next, a fraction of these cold atoms, on the order of $10^5$–$10^6$, is loaded into a red-detuned optical dipole trap (ODT) \cite{2000FresePRL,2008ZirbelPRL,1996LeePRL} at 1064\,nm. The inhomogeneous intensity profile of the ODT further reduces their temperature to a few microkelvin. By adiabatically transporting the ODT beam via a frequency-shifted optical conveyor, we transfer the atoms into the mode volume of a high-finesse Fabry–P\'{e}rot cavity.

Finally, we apply two phase-locked coupling lasers on the $\Lambda$-type subsystem formed by the $5P_{1/2}$, $5P_{3/2}$, and $6S_{1/2}$ levels. These two coherent fields drive the ground state $5S_{1/2}$ into a long-lived dark-state polariton, thereby realizing the desired QB.

\section{Conclusion}
\label{sec:Conclusion}

In summary, we introduce EIT into a four-level atom and transform it into an effectively two-level system consisting of the ground state and the dark state. We then replace the two-level atoms in the QB with these quasi-two-level atoms. Three distinct models are developed: one featuring a single atom and a single photon within a cavity, another with multiple atoms and a single photon, and the third with multiple atoms and multiple photons. We also consider the impact of the dipole-dipole interaction on the ergotropy. Finally, we explore the possibility of using pulses to excite our QB. We investigate the evolution of the QB's energy and ergotropy in these models and find that the introduction of EIT significantly suppresses the decay of both energy and ergotropy compared to scenarios without EIT. Under the specified parameter conditions, the decay rate of the system is reduced by two orders of magnitude upon incorporating EIT. Moreover, by adjusting the coupling frequency $\omega_c$, the ergotropy of the system can be further enhanced. This study demonstrates that the introduction of EIT can significantly extend the lifetime of QBs. Additionally, the energy transfer efficiency during the charging process is improved. These advancements not only enhance the performance of QBs but also pave the way for their broader application in various technological fields.

\section{Acknowledgments}
\label{sec:Acknowledgments}
This work was supported by the Innovation Program for Quantum Science and Technology under Grant No.~2023ZD0300200, the National Natural Science Foundation of China under Grant No.~62461160263, the Beijing Natural Science Foundation under Grant No.~1202017, and the Beijing Normal University under Grant No.~2022129.

\appendix

\section{Diagonization of $H_{0}$}
\label{sec:AppendixA}

In this appendix, we will diagonalize $H_{0}$ by the perturbation theory. According to the Schr\"{o}dinger equation $H|x\rangle=x|x\rangle$, we can obtain the eigen energies {$x$'s} by the following equation, i.e.,
\begin{equation}
x\left[x^{2}-(\omega_{1}+\omega_{2})x+\omega_{1}\omega_{2}-\Omega^{2}\right]+\Omega_{1}^{2}\omega_{2}=0,\label{eq:x}
\end{equation}
where
\begin{equation}
\Omega^{2}=\Omega_{1}^{2}+\Omega_{2}^{2}.
\end{equation}

When $\omega_{2}$ is small, to the first order of $\omega_{2}$, the
solutions can be written as
\begin{equation}
x_{j}\simeq x_{0j}+A_{j}\omega_{2}.
\end{equation}
According to the perturbation theory, the zero-order terms are given
by
\begin{equation}
x_{0j}\left[x_{0j}^{2}-(\omega_{1}+\omega_{2})x_{0j}+\omega_{1}\omega_{2}-\Omega^{2}\right]=0,\label{eq:x0}
\end{equation}
where
\begin{align}
x_{01} & =0,\\
x_{02} & =\omega_{+},\\
x_{03} & =\omega_{-},\\
\omega_{\pm} & =\frac{1}{2}\left[(\omega_{1}+\omega_{2})\pm\sqrt{(\omega_{1}-\omega_{2})^{2}+4\Omega^{2}}\right]
\end{align}

After some algebra, we can rewrite Eq.~(\ref{eq:x}) as
\begin{equation}
x(x-\omega_{+})(x-\omega_{-})+\Omega_{1}^{2}\omega_{2}=0.\label{eq:xp}
\end{equation}
By substituting $x_{1}^{\prime}=A_{1}\omega_{2}$ into the above equation,
we have
\begin{equation}
A_{1}\omega_{2}(A_{1}\omega_{2}-\omega_{+})(A_{1}\omega_{2}-\omega_{-})+\Omega_{1}^{2}\omega_{2}=0.
\end{equation}
Ignoring higher-order terms of $\omega_{2}$, we can obtain
\begin{equation}
A_{1}=-\frac{\Omega_{1}^{2}}{\omega_{+}\omega_{-}}\simeq\frac{\Omega_{1}^{2}}{\Omega^{2}}.
\end{equation}
By repeating the above procedure, we have
\begin{align}
A_{2} & =-\frac{\Omega_{1}^{2}}{\omega_{+}(\omega_{+}-\omega_{-})}\simeq-\frac{\Omega_{1}^{2}}{2\Omega^{2}},\\
A_{3} & =\frac{\Omega_{1}^{2}}{\omega_{-}(\omega_{+}-\omega_{-})}\simeq-\frac{\Omega_{1}^{2}}{2\Omega^{2}}.
\end{align}

When $\omega_{1},\omega_{2}\ll\Omega_{1},\Omega_{2}$, to the first-order
terms of {$\omega_{1}$ and} $\omega_{2}$, we can obtain
\begin{align}
\omega_{\pm} &  \simeq\frac{1}{2}[(\omega_{1}+\omega_{2})\pm2\Omega].
\end{align}
In all, the eigen energies are respectively
\begin{align}
x_{1}^{\prime} & \simeq\frac{\Omega_{1}^{2}}{\Omega^{2}}\omega_{2},\\
x_{2} & \simeq\Omega+\frac{1}{2}\biggl(\omega_{1}+\frac{\Omega_{2}^{2}}{\Omega^{2}}\omega_{2}\biggr),\\
x_{3} &  \simeq-\Omega+\frac{1}{2}\biggl(\omega_{1}+\frac{\Omega_{2}^{2}}{\Omega^{2}}\omega_{2}\biggr).
\end{align}
Correspondingly, the eigen states are
\begin{align}
|E_{i}\rangle=  \frac{1}{N_{i}}\big\{&\big[(x_{i}-\omega_{1})(x_{i}-\omega_{2})-\Omega_{2}^{2}\big]|g\rangle\\
 & +\Omega_{1}(x_{i}-\omega_{2})|e\rangle+\Omega_{1}\Omega_{2}|m\rangle\big\},
\end{align}
where {the normalization constants $N_{i}$'s are given by
\begin{align}
N_{i}^{2}=&\left|(x_{i}-\omega_{1})(x_{i}-\omega_{2})-\Omega_{2}^{2}\right|^{2}+\left|\Omega_{1}(x_{i}-\omega_{2})\right|^{2}\nonumber\\
&+\left|\Omega_{1}\Omega_{2}\right|^{2}.
\end{align}}
To the zero-order terms of {$\omega_{1}$ and} $\omega_{2}$, we have
\begin{align}
|E_{1}\rangle & \simeq\frac{\Omega_{2}}{N_{1}}(-\Omega_{2}|d\rangle+\Omega_{1}|m\rangle)\\
|E_{2}\rangle & \simeq\frac{\Omega_{1}}{N_{2}}(\Omega_{1}|d\rangle+\Omega|e\rangle+\Omega_{2}|m\rangle),\\
|E_{3}\rangle & \simeq\frac{\Omega_{1}}{N_{3}}(\Omega_{1}|d\rangle-\Omega|e\rangle+\Omega_{2}|m\rangle),
\end{align}
where $|E_{1}\rangle$ is the dark state because it is a superposition
of $|d\rangle$ and $|m\rangle${. The} other two eigen states 
$|E_{2}\rangle$ and $|E_{3}\rangle$ are referred to as the bright
states, because in {addition} to $|m\rangle$ they also contain $|e\rangle$,
which suffer{s} more from decoherence. The decoherence will cause the
aging of the QB. Therefore, we introduce $|E_{1}\rangle$ to the QB
in order to suppress the decoherence.

\section{Reduced Density Matrix of QB}
\label{sec:AppendixB}

In this appendix, we derive the maximum extractable energy. According
to Eqs.~(\ref{eq:19})-(\ref{eq:20}), $|\psi_{n}^{+}\rangle$ and
$|\psi_{n}^{-}\rangle$ can be written {in terms of the basis $|n\rangle|E_{1}\rangle$ and $|n+1\rangle|g\rangle$} as
\begin{align}
|\psi_{n}^{\pm}\rangle & =\left(\begin{array}{c}
c_{\pm}\\
d_{\pm}
\end{array}\right),\label{eq:19-1}
\end{align}
where
\begin{align}
c_{\pm} & =\frac{J}{\sqrt{\left(x_{1}^{\prime}+n\omega-\lambda_{n}^{\pm}\right)^{2}+J^{2}}},\label{eq:19-1-1}\\
d_{\pm} & =\frac{\lambda_{n}^{\pm}-x_{1}^{\prime}-n\omega}{\sqrt{\left(x_{1}^{\prime}+n\omega-\lambda_{n}^{\pm}\right)^{2}+J^{2}}}.
\end{align}

{In the basis $|n\rangle|E_{1}\rangle$ and $|n+1\rangle|g\rangle$,}
the time-evolution operator
\begin{equation}
U(t)=e^{-iHt}=\sum_{\alpha=\pm}e^{-i\lambda_{n}^{\alpha}t}|\psi_{n}^{\alpha}\rangle\langle\psi_{n}^{\alpha}\vert,
\end{equation}
can be rewritten as
\begin{align}
U^{\prime}(t) & =PU(t)P^{-1}\nonumber \\
 & =\left(\begin{array}{cc}
\frac{c_{-}d_{+}e^{-i\lambda_{n}^{-}t}-c_{+}d_{-}e^{-i\lambda_{n}^{+}t}}{c_{-}d_{+}-c_{+}d_{-}} & \frac{c_{+}c_{-}(e^{-i\lambda_{n}^{+}t}-e^{-i\lambda_{n}^{-}t})}{c_{-}d_{+}-c_{+}d_{-}}\\
\frac{d_{+}d_{-}(e^{-i\lambda_{n}^{-}t}-e^{-i\lambda_{n}^{+}t})}{c_{-}d_{+}-c_{+}d_{-}} & \frac{c_{-}d_{+}e^{-i\lambda_{n}^{+}t}-c_{+}d_{-}e^{-i\lambda_{n}^{-}t}}{c_{-}d_{+}-c_{+}d_{-}}
\end{array}\right),
\end{align}
where 
\begin{eqnarray}
P & = & \left(\begin{array}{cc}
c_{+} & c_{-}\\
d_{+} & d_{-}
\end{array}\right).
\end{eqnarray}

Initially, the system is in the state $|\psi_{0}\rangle=\alpha|n\rangle|E_{1}\rangle+\beta|n+1\rangle|g\rangle$.
The density matrix $\rho(t)=U^{\prime}(t)\rho(0)U^{\prime\dagger}(t)$ at time
$t$ reads
\begin{equation}
\rho(t)=\frac{1}{|c_{2}d_{1}-c_{1}d_{2}|^{2}}\begin{pmatrix}M_{11} & M_{12}\\
M_{21} & M_{22}
\end{pmatrix},
\end{equation}
where\begin{widetext}
\begin{eqnarray}
M_{11} & \!\!=\!\! & \left(c_{+}^{*}d_{-}^{*}e^{i\lambda_{n}^{+}t}-c_{-}^{*}d_{+}^{*}e^{i\lambda_{n}^{-}t}\right)\times\left[(c_{+}d_{-}e^{-i\lambda_{n}^{+}t}-c_{-}d_{+}e^{-i\lambda_{n}^{-}t})|\alpha|^{2}+c_{+}c_{-}(e^{-i\lambda_{n}^{+}t}-e^{-i\lambda_{n}^{-}t})\beta\alpha^{*}\right]\nonumber \\
 & \!\!\!\! & +c_{+}^{*}c_{-}^{*}(e^{i\lambda_{n}^{+}t}-e^{i\lambda_{n}^{-}t})\times\left[c_{+}c_{-}(e^{-i\lambda_{n}^{+}t}-e^{-i\lambda_{n}^{-}t})|\beta|^{2}+(-c_{+}d_{-}e^{-i\lambda_{n}^{+}t}+c_{-}d_{+}e^{-i\lambda_{n}^{-}t})\alpha\beta^{*}\right],\\
M_{12} & \!\!=\!\! & d_{+}^{*}d_{-}^{*}(e^{i\lambda_{n}^{+}t}-e^{i\lambda_{n}^{-}t})\times\left[(c_{+}d_{-}e^{-i\lambda_{n}^{+}t}-c_{-}d_{+}e^{-i\lambda_{n}^{-}t})|\alpha|^{2}+c_{+}c_{-}(e^{-i\lambda_{n}^{+}t}-e^{-i\lambda_{n}^{-}t})\beta\alpha^{*}\right]\nonumber \\
 & \!\!\!\! & +\left(c_{-}^{*}d_{+}^{*}e^{i\lambda_{n}^{+}t}-c_{+}^{*}d_{-}^{*}e^{i\lambda_{n}^{-}t}\right)\times\left[c_{+}c_{-}(e^{-i\lambda_{n}^{+}t}-e^{-i\lambda_{n}^{-}t})|\beta|^{2}+(-c_{+}d_{-}e^{-i\lambda_{n}^{+}t}+c_{-}d_{+}e^{-i\lambda_{n}^{-}t})\alpha\beta^{*}\right],\\
M_{21} & \!\!=\!\! & \left(c_{+}^{*}d_{-}^{*}e^{i\lambda_{n}^{+}t}-c_{-}^{*}d_{+}^{*}e^{i\lambda_{n}^{-}t}\right)\times\left[d_{+}d_{-}(e^{-i\lambda_{n}^{+}t}-e^{-i\lambda_{n}^{-}t})|\alpha|^{2}+(-c_{-}d_{+}e^{-i\lambda_{n}^{+}t}+c_{+}d_{-}e^{-i\lambda_{n}^{-}t})\beta\alpha^{*}\right]\nonumber \\
 & \!\!\!\! & +c_{+}^{*}c_{-}^{*}(e^{i\lambda_{n}^{+}t}-e^{i\lambda_{n}^{-}t})\times\left[(c_{-}d_{+}e^{-i\lambda_{n}^{+}t}-c_{+}d_{-}e^{-i\lambda_{n}^{-}t})|\beta|^{2}+d_{+}d_{-}(-e^{-i\lambda_{n}^{+}t}+e^{-i\lambda_{n}^{-}t})\alpha\beta^{*}\right],\\
M_{22} & \!\!=\!\! & d_{+}^{*}d_{-}^{*}(e^{i\lambda_{n}^{+}t}-e^{i\lambda_{n}^{-}t})\times\left[d_{+}d_{-}(e^{-i\lambda_{n}^{+}t}-e^{-i\lambda_{n}^{-}t})|\alpha|^{2}+(-c_{-}d_{+}e^{-i\lambda_{n}^{+}t}+c_{+}d_{-}e^{-i\lambda_{n}^{-}t})\beta\alpha^{*}\right]\nonumber \\
 & \!\!\!\! & +\left(c_{-}^{*}d_{+}^{*}e^{i\lambda_{n}^{+}t}-c_{+}^{*}d_{-}^{*}e^{i\lambda_{n}^{-}t}\right)\times\left[(c_{-}d_{+}e^{-i\lambda_{n}^{+}t}-c_{+}d_{-}e^{-i\lambda_{n}^{-}t})|\beta|^{2}+d_{+}d_{-}(-e^{-i\lambda_{n}^{+}t}+e^{-i\lambda_{n}^{-}t})\alpha\beta^{*}\right].
\end{eqnarray}
\end{widetext}
Thus, the reduced density matrix of the QB is 
\begin{align} \rho_{B}(t) & =\frac{1}{|c_{2}d_{1}-c_{1}d_{2}|^{2}}\begin{pmatrix}M_{11} & 0\\0 & M_{22}\end{pmatrix}.
\end{align}
By substituting the above equation into $W(t)=\mathrm{Tr}[\rho_{B}(t)H_{B}]-\mathrm{Tr}[\tilde{\rho}_{B}(t)H_{B}]$, we can obtain the time evolution of the system's ergodicity.

\section{Time Evolution by Wei-Norman Algebra}
\label{sec:AppendixC}

In this appendix, we will unravel the time evolution by Wei-Norman algebra.

Since
\begin{eqnarray}
[ab^{\dagger},b^{\dagger}b-a^{\dagger}a] & = & -2ab^{\dagger},\\{}
[a^{\dagger}b,b^{\dagger}b-a^{\dagger}a] & = & 2a^{\dagger}b,\\{}
[ab^{\dagger},a^{\dagger}b] & = & b^{\dagger}b-a^{\dagger}a,
\end{eqnarray}
the set of operators $\{ab^{\dagger},a^{\dagger}b,b^{\dagger}b-a^{\dagger}a\}=\{H_{1},H_{2},H_{3}\}$
forms a complete set. In order to solve the time{-}evolution {operator}
$U(t)=\exp(-iHt)$, we can use the Wei-Norman algebra \cite{Wei1963JMP,Xu2011PRA}.
The Hamiltonian $H$ can be rewritten in terms of the above operators
as $H=f_{1}H_{1}+f_{2}H_{2}+f_{3}H_{3}$. Thus, the time-evolution
operator {in the interaction picture} can be given as the product of a series of exponential operators, i.e.,
\begin{equation}
U_{I}(t)=\prod_{j=1}^{3}\exp[g_{j}(t)H_{j}],\label{eq:UI}
\end{equation}
where $g_{i}(t)$ is a function of time and $H_{i}$'s are the generators
of the Wei-Norman algebra. In the interaction picture, the time-evolution
operator is determined by
\begin{eqnarray}
i\frac{dU_{I}(t)}{dt} & = & H_{I}U_{I}(t).
\end{eqnarray}
By substituting Eq.~(\ref{eq:UI}) into the above equation, we have
\begin{eqnarray}
H_{I} & = & i(\dot{g_{1}}H_{1}+e^{g_{1}H_{1}}\dot{g_{2}}H_{2}e^{-g_{1}H_{1}}\nonumber \\
 &  & +e^{g_{1}H_{1}}e^{g_{2}H_{2}}\dot{g_{3}}H_{3}e^{-g_{2}H_{2}}e^{-g_{1}H_{1}}).
\end{eqnarray}
{Using} the Baker-Campbell-Hausdorff formula \cite{Sakurai1994}, we
can obtain
\begin{eqnarray}
e^{g_{1}H_{1}}\dot{g}_{2}H_{2}e^{-g_{1}H_{1}} & = & \dot{g_{2}}[a^{\dagger}b+g_{1}(b^{\dagger}b-a^{\dagger}a)-g_{1}^{2}ab^{\dagger}].\nonumber \\
\label{eq:16}
\end{eqnarray}
By repeating the above procedure, we have
\begin{align}
 &\dot{g}_{3}[2g_{2}a^{\dagger}b-(2g_{1}+2g_{1}^{2}g_{2})ab^{\dagger}+(1+2g_{1}g_{2})(b^{\dagger}b-a^{\dagger}a)]\nonumber \\
 &=e^{g_{1}H_{1}}e^{g_{2}H_{2}}\dot{g}_{3}H_{3}e^{-g_{2}H_{2}}e^{-g_{1}H_{1}}.\label{eq:17}
\end{align}
By substituting Eqs.~(\ref{eq:16}) and (\ref{eq:17}) into $H_{I}$,
we can obtain
\begin{eqnarray}
H_{I} & = & i\left\{ (\dot{g}_{2}+2\dot{g}_{3}g_{2})a^{\dagger}b\right.\nonumber \\
 &  & +\left[\dot{g}_{1}-\dot{g}_{2}g_{1}^{2}-\dot{g}_{3}(2g_{1}+2g_{1}^{2}g_{2})\right]ab^{\dagger}\nonumber \\
 &  & \left.+\left[\dot{g}_{2}g_{1}+\dot{g}_{3}(1+2g_{1}g_{2})\right](b^{\dagger}b-a^{\dagger}a)\right\} .
\end{eqnarray}
By comparing the above equation with $H_{I}=J\sqrt{N}(ab^{\dagger}+a^{\dagger}b)$,
we have
\begin{eqnarray}
\dot{g}_{1}-\dot{g}_{2}g_{1}^{2}-\dot{g}_{3}(2g_{1}+2g_{1}^{2}g_{2}) & = & -iJ_{N},\\
\dot{g}_{2}+2\dot{g}_{3}g_{2} & = & -iJ_{N},\\
\dot{g}_{2}g_{1}+\dot{g}_{3}(1+2g_{1}g_{2}) & = & 0.
\end{eqnarray}
Assuming the initial condition $g_{1}(0)=g_{2}(0)=g_{3}(0)=0$, we
can obtain
\begin{eqnarray}
g_{1} & = & -i\tan (J_{N}t),\\
g_{2} & = & -\frac{i}{2}\sin(2J_{N}t),\\
g_{3} & = & -\ln[\cos(J_{N}t)].
\end{eqnarray}
Therefore, the time-evolution operator reads
\begin{eqnarray}
U(t) & = & U_{0}(t)U_{I}(t),\\
U_{0}(t) & = & \exp\left[-i\omega_{0}t(a^{\dagger}a+b^{\dagger}b)\right],\\
U_{I}(t) & = & e^{g_{1}ab^{\dagger}}e^{g_{2}a^{\dagger}b}e^{g_{3}(b^{\dagger}b-a^{\dagger}a)}.
\end{eqnarray}





%

\end{document}